\documentclass[12pt]{article}
\usepackage{amsfonts}
\usepackage[slantedGreek]{mathpazo}
\usepackage{amssymb}
\usepackage{graphicx}
\graphicspath{{fig/}}
\usepackage{color}
\usepackage{amsmath}
\usepackage{amsthm}
\usepackage{geometry}[margin=0.8in]
\usepackage{natbib}

\def\beq{\begin{equation}}
\def\be{\beq}

\def\eea{\end{eqnarray}} \def\eeaa{\end{eqnarray*}}
\def\eeq{\end{equation}}  
   
\def\ee{\eeq}

\newtheorem{theorem}{Theorem}

\begin{document}

\title{$MC^2$  Mixed Integer and Linear Programming}
\author{	
    \makebox[.4\linewidth]{Nick Polson}\\\textit{Booth School of Business}\\\textit{University of Chicago}\\\and 
    \makebox[.4\linewidth]{Vadim Sokolov\footnote{Nick Polson is at Chicago Booth: ngp@chicagobooth.edu. Vadim Sokolov is Associate Professor at Volgenau School of Engineering, George Mason University, USA: vsokolov@gmu.edu.}}\\\textit{Department of Systems Engineering}\\\textit{and Operations Research}\\\textit{George Mason University}
}
\date{First Draft: December 25, 2024\\This Draft: \today}

\maketitle
\begin{abstract}
\noindent  
In this paper, we design $MC^2$ algorithms for Mixed Integer and Linear Programming. By expressing a constrained optimisation as one of simulation from a Boltzmann distribution, we reformulate integer and linear programming as Monte Carlo optimisation problems. The key insight is that solving these optimisation problems requires the ability to simulate from truncated distributions, namely multivariate exponentials and Gaussians. Efficient simulation can be achieved using the algorithms of Kent and Davis. We demonstrate our methodology on portfolio optimisation and the classical farmer problem from stochastic programming. Finally, we conclude with directions for future research.
\end{abstract}


\newpage 

\section{Introduction}
The connection between optimisation and simulation has a rich history in the operations research and statistics literature. Following the seminal work of \cite{pincus1968closed,pincus1970monte}, who established the theoretical foundations for using Monte Carlo methods to solve optimisation problems, and \cite{kirkpatrick1983optimization}, who introduced simulated annealing as a stochastic optimisation technique, there has been sustained interest in simulation-based approaches to combinatorial and continuous optimisation. \cite{geman1990boundary} extended these ideas to constrained optimisation by introducing soft constraint penalties. In parallel, \cite{besag1974spatial} used the Boltzmann distribution and an iterated conditional mode estimation procedure in image processing. \cite{gelfand1992bayesian} considered the general problem of simulation from truncated distributions, which is central to our approach.

In the foundational work, Pincus considered the problem of finding $\max_x f(x)$ subject to $g(x) = 0$. Geman proposed to solve this via an exponentially tilted sampling problem by defining
\[
\pi_{\kappa,\lambda}(x) \propto \exp \{-\kappa (f(x) + \lambda_\kappa g(x))\}
\]
where $\kappa$ and $\lambda_\kappa$ are annealing parameters. This formulation is closely related to the Lagrange multiplier approach, and the parameter $\lambda_\kappa$ plays a role analogous to dual variables in constrained optimisation.

That is, take the Pincus density $\pi_\kappa(x) \propto \exp \{-\kappa f(x)\}$ and replace the constraint $g(x) = 0$ by a ``soft'' exponential tilting $\exp \{-\lambda_\kappa g(x)\}$.

The advantage of this approach is that it can handle nonlinear constraints $g(x)$. Moreover, EM-style algorithms can be constructed by using a latent variable scheme for both densities. If we augment with latents $(\lambda, \omega)$ then we can write
\begin{align*}
\pi_{\kappa,\lambda}(x) &\propto \exp \{-\kappa (f(x) + \lambda_\kappa g(x))\} \\
&= e^{ax} \mathbb{E}_\omega \left\{ e^{-\frac{x^2}{2}\omega} \right\} e^{bx} \mathbb{E}_\lambda \left\{ e^{-\frac{x^2}{2}\lambda} \right\}
\end{align*}
where the parameters $(a, b)$ depend on $(\kappa, \lambda_\kappa)$ and the functional forms of $(f(x), g(x))$.

This latent variable representation enables the construction of an EM algorithm that is quadratic in $x$ given $(\omega, \lambda)$. The conditional moments $\mathbb{E}(\omega|x)$ and $\mathbb{E}(\lambda|x)$ are available in terms of derivatives of $f$ and $g$.

A number of simulation techniques can be employed, including nested sampling, vertical likelihood sampling, and MCMC methods. For simulated annealing tasks, standard MCMC approaches can perform poorly due to the difficulty of traversing energy barriers. 
In the integer programming case, where the simulation is supported on a discrete set, we recommend the use of a stochastic water-filling method to reduce the number of particles in the simulation set while retaining the required Monte Carlo properties. The use of auxiliary variables (slice variables) is particularly useful in multi-modal cases, as they enable the sampler to escape local optima more easily.

MCMC methods can achieve polynomial-time complexity for certain combinatorial problems: \cite{polson2024counting} demonstrated this for the $N$-Queens problem, and \cite{kannan1987minkowskis} established similar results for integer programming under appropriate conditions.

The central idea of our approach is straightforward. Suppose that we wish to solve the optimisation problem
$$
x^\ast = {\rm arg} \max_{ x \in \mathcal{X} }  \; f ( x ) \; \;  {\rm subject \; to}  \; \;  g(x) < 0 
$$
Then we replace this with the problem of finding the mode of the Boltzmann distribution
$$
\pi_{\kappa} ( x ) = \frac{1}{Z_\kappa}  \exp \left ( - \kappa f( x) \right )  \mathbb{I}_{ g(x) < 0 } .
$$
This can be achieved in an annealed fashion. 

In some formulations, the dual problem replaces a hard equality constraint with an inequality constraint. Both types of constraints can be handled naturally within our Monte Carlo framework.

Our methodology is particularly well-suited for one-stage and two-stage stochastic programming problems with uncertainty. In these settings, simulation-based methods are essential because the objective function involves expectations that are not available in closed form \citep{ekin2017augmented,ekin2014augmented}.

The key computational challenge is that one must simultaneously calculate an expectation and optimise. A brute-force Monte Carlo approach first estimates the expectation and then optimises, but this can be highly inefficient, particularly in rare event simulation problems. Our MCMC approach allows both tasks to be performed simultaneously.

Related work by \cite{glynn2010how} uses splitting methods to count the number of solutions to integer programming problems, which shares the spirit of our simulation-based approach.

The rest of the paper is organised as follows. In the remainder of this section, we establish the theoretical foundations for optimisation via Boltzmann distributions and discuss efficient sampling from discrete distributions. Section~2 develops the $MC^2$ methodology for integer and linear programming problems, using the annealed stochastic optimisation approach of Pincus. Section~3 presents stochastic sampling techniques for truncated multivariate exponentials and normals, including the stochastic water-filling representation for particle economisation. Section~4 demonstrates the methodology on the classical farmer problem from stochastic programming. We conclude with a discussion of future research directions. Technical details on one-stage and two-stage stochastic programming formulations, along with slice sampling methods, are provided in an Appendix.

\subsection{Optimisation with Constraints}

The general problem we address is to find $ {\rm arg min}_x f(x)$ where $ x \in \mathcal{X} $ for some domain $ \mathcal{X} $ with a given objective function, $f(x)$, which is assumed to have a finite minimum attainable in $ \mathcal{X} $ that
may exhibit multi-modality. We define the set of minima as
$$
\mathcal{X}_{min} = \{  x \in \mathcal{X} : f( x) = \min_y f(y) \} \; .
$$
We will develop a simulation-based method to find $\mathcal{X}_{min}   $ by exploiting a well-known duality between the problem above and
that of  finding the modes of the Boltzmann
distribution with energy potential, $f(x)$, and density defined by
$$
\pi_\kappa (x) = \exp \left \{ - \kappa f(x) \right \} / Z_\kappa \; \; {\rm for} \; \; x \in \mathcal{X} \; .
$$
This density is indexed by a  ``temperature'' parameter, $\kappa$, and
$ Z_\kappa = \int_{\mathcal{X}}  \exp \left \{ - \kappa f(x) \right \} d x $ is 
the normalisation constant, or partition function. Like simulated tempering, we have to perform a sensitivity analysis with respect to
the cooling parameter $\kappa$ by specifying an initial set of values.
A number of methods, e.g. Wang-Landau algorithm and its generalisations and multi-canonical sampling \cite{berg1991multicanonical},
place pseudo-prior weights over $\kappa$, denoted by $p(\kappa)$, and use the mixture distribution $ \sum_{\kappa} p(\kappa) \pi_\kappa (x) $
to traverse the multi-modal distribution. We allow for this possibility in our framework, although as we show empirically, the use of slice
variables greatly enhances the ability of the Markov chain to traverse low and high energy states of the sample space.

We will use Markov chain MC simulation methods to sample from this
possibly high dimensional joint distribution. One advantage of this approach is that they  will not be require explicit knowledge of $Z_\kappa $.
There are, however, many possible Markov transition dynamics that have the appropriate equilibrium distribution.
Here we propose a new method based on exponential slice sampling.

The limiting cases, $ \kappa \in \{ 0 , \infty \} $ both lead to a uniform measure but on different sets.
For $ \kappa =0 $, we have $ \pi_0 (x) $ as the uniform measure on $ \mathcal{X} $. For $ \kappa \rightarrow \infty $,
we have a uniform measure on the desired set $ \mathcal{X}_{min} $. Specifically, we have
$$
\lim_{ \kappa \rightarrow \infty } \pi_\kappa (x) = \pi_{\infty} ( x ) = | \mathcal{X}_{min} |^{-1} \delta_{ \mathcal{X}_{min} } (x)
$$
where $ \delta_x$ denotes Dirac measure.

The asymptotic in $\kappa $ is the basis of simulated annealing \cite{kirkpatrick1983optimization,vanlaarhoven1987simulated,aarts1988simulated,muller2004optimal} which uses a schedule of parameter values $ \kappa^{(g)} $ that increases with the length of the Markov chain simulation in an appropriate fashion \cite{gidas1985nonstationary}.
Other approaches include simulated tempering which uses a random walk on a set of temperature values $ 0 < \kappa_1 < \ldots < \kappa_m $ rather than increasing $\kappa$ on a schedule, equi-energy sampling \cite{kou2006equienergya}, evolutionary MCMC and the Wang-Landau algorithm..

There are a number of ways of computing the mode: for example, Markov chain
simulation methods. For a fixed $\kappa$, consider running Markov chain 
$$ 
X^{(0)} , X^{(1)} , \ldots , X^{(G)} , \ldots 
$$
with a transition kernel defined so as to have the appropriate equilibrium distribution $ \pi_\kappa (x)$. Then under mild Harris recurrence conditions,
given any starting point,
$$ \lim_{ G \rightarrow \infty} \mathbb{P} \left ( X^{(G)} \in A | X^{(0)}= x \right ) = \pi_\kappa (A ),
$$
for any Borel sets $A$. The main issue is determining which Markov chain to use. We argue for the use of slice sampling methods
for a set of temperature values defined in a given set in a similar fashion to simulated tempering.

In seminal work, \cite{pincus1968closed,pincus1970monte} proposed to directly use a Metropolis algorithm to simulate from
the Gibbs distribution, then after discarding a burn-in period, to use the ergodic mean along the chain $ \frac{1}{G} \sum_{g=1}^G X^{(g)} $ as an estimate of the mode.
In the uni-modal case, as $ \kappa \rightarrow \infty $, this will find the mode. The simulated annealing literature
also includes $ \kappa $ as a control parameter in the Markov chain, which indexes the transition kernel with $ \kappa^{(g)} $.
This now makes the simulation procedure a time-inhomogeneous chain and conditions on the choice of schedule $ \kappa^{(g)} $ are
required to guarantee convergence to the mode, see \cite{gidas1985nonstationary,geman1990boundary} for further discussion.

We note that even in the multimodal case,  a suitably defined Markov chain will converge
to a uniform measure over the modes. By monitoring the output of the chain, in principle we can find all the modes.
On the theoretical side, the issue is where the constructed Markov chain converges in polynomial time so that the conductance \cite{polson1996convergence}
of moving from mode to mode for a large $\kappa$ is high enough. Clearly, in hard multi-modal cases this is unlikely to be the case, as
the surfaces typically have witches' hat spikes where it is known that, even though there is geometric convergence of a simple Gibbs sampler, it is not polynomial.

Only difference is that for integer programming we are sampling from a discrete set. There are a number of methods to stochastically sample discrete distributions \cite{bertsimas2005optimization}.


When the optimisation problem has a discrete feasible region, we require efficient methods for sampling from discrete probability distributions. \cite{fearnhead2003line} proposed an $O(N)$ algorithm for exact simulation from discrete distributions that avoids the $O(N \log N)$ complexity of methods based on sorting or balanced tree structures. Their approach uses a clever partitioning scheme that allows direct sampling in linear time.

An alternative approach is to use nested sampling, as developed in the context of Bayesian computation by \cite{skilling2006nested}. This method constructs a sequence of nested constraint sets and can be particularly effective when the target distribution concentrates on a small subset of the discrete space. The combination of these techniques with our MCMC framework provides a complete toolkit for discrete optimisation problems.

We defer the detailed discussion of one-stage and two-stage stochastic programming formulations to the Appendix. 

\section{$MC^2$ for Integer and Linear Programming}

We now develop the $MC^2$ methodology for linear programming problems. Consider the LP problem of calculating
$$
G(z) = \max_\pi \left \{ \pi^\prime z : W^\prime \pi \leq q \right \}
$$
Following Pincus, we consider the annealed distribution
$$
p_\kappa ( \pi | z , q , W ) = \exp \left ( - \kappa \pi z \right ) \mathbb{I} \left ( W^\prime \pi \leq q \right ) / Z_\kappa
$$
where $ Z_\kappa = \int_{  W^\prime \pi \leq q }\exp \left ( - \kappa \pi z \right ) d \pi $ is an appropriate normalisation constant.

Then as $ \kappa \rightarrow \infty $, this distribution tends to a Dirac measure on the solution $ \pi^\star $ of the LP.
Simulation from $ p_\kappa ( \pi) $, however, requires a method for dealing with truncated multivariate exponential distributions.

We can then simulate a Markov chain and obtain draws $ \pi^{(h)} $ for $ h =1, \ldots , H$ and
estimate the value function as a Monte Carlo Rao-Blackwellised average $\hat{G} ( z ) = \frac{1}{H} \sum_{i=1}^H \pi^{(g)} z $,
an average of piecewise linear functions.

\subsection{Example: \cite{pincus1968closed}}

Consider the problem with $ b ,t >0 $
$$
 \max c_1 x_1 + c_2 x_2 \; \; {\rm  subject \; to} \; \;
 x_1 + b x_2 = t
$$
where we also have the constraint $ x_1 , x_2 >0 $.

An alternative approach is to substitute out a variable. For instance, using $x_1 = t - b x_2$ introduces a nonlinear term $(t - b x_2)_+$ to enforce the positivity constraint on $x_1$. We can still use the Pincus marginal annealed distribution with this formulation. Furthermore, data augmentation or EM-style algorithms can handle the $\exp(\kappa \max(\cdot))$ term that arises from the positivity constraint by introducing appropriate latent variables.

The solution is given by
$$
x_1^{\star} = t \; {\rm if} \;  c_1 > 0 , b c_1 <  c_2
$$
and otherwise $ x_1^{\star} = 0 $. For the other variable
$$
x_2^{\star} = \frac{t}{b} \; {\rm if} \;  c_2 > 0 , b c_1 <  c_2
$$
and otherwise $ x_2^{\star} = 0 $.

We can add two slack variables for the greater than zero constraints and solve
$ \max c^T x $ subject to $ A x = b $ where
$ x^T =  ( x_1 , x_2 , x_3 , x_4 ) $ and
$$
A = \left ( \begin{array}{cccc}
1 & b & 0 & 0 \\
1 & 0 & 1 & 0 \\
0 & 1 & 0 & 1 \\
\end{array}\right )
\; \; b =  \left ( \begin{array}{c}
t \\
1 \\
1 \\
\end{array} \right )
$$
Moreover,by
Duality (see \cite{birge1997introduction}, p. 73)
$$
\max \{ c^T x \; | \; A x = b \} \; \; \; {\rm  equivalent \; to} \; \;  \;
\min \{ \pi^T b \; | \; \pi^T A \geq c^T \}
$$
where
$ \pi$ are called dual variables. The optimal solutions satisfy $ c^T x^{\star} =  ( \pi^{\star} )^T b $

The Dual variables define a Boltzmann-Gibbs distribution for $ \pi =  ( \pi_1 , \pi_2 , \pi_3 )$ defined by
$$
p_{ \kappa } ( \pi ) \propto e^{ - \kappa ( t \pi_1 + \pi_2 + \pi_3 ) } \mathbb{I} \left ( \pi^T A \geq c^T \right )
$$
The Gibbs sampler iterates through the following truncated exponential conditionals:
\begin{itemize}
\item $ \pi_1 | \pi_2 , \pi_3 \sim  e^{ - \kappa \pi_1 t } \mathbb{I} \left ( \max ( c_1 - \pi_2 ,
\frac{ c_2 - \pi_3 }{b} ) , \infty \right ) $
\item $ \pi_2 | \pi_1 , \pi_3 \sim  e^{ - \kappa \pi_2 } \mathbb{I} \left ( \max ( c_1 - \pi_1 , 0 ) , \infty \right ) $
\item $ \pi_3 | \pi_1 , \pi_2 \sim  e^{ - \kappa \pi_3 } \mathbb{I} \left ( \max ( c_2 - b \pi_1 , 0 ) , \infty \right )$
\end{itemize}

Depending on the sign of $ c_2 - b c_1 $ the conditional for $ \pi_3$ will converge to the desired two solutions.
If $ c_2 < b c_1 $, then $ \pi_3 \approx 0 $ and then
$ \pi_1 \approx c_1 $ and $ \pi_2 \approx 0 $. Hence $ x^{\star} $.

If $ c_2 > b c_1 $, then $ \pi_3 \approx  c_2 - b \pi_1 $.
Then $ \frac{ c_2 - \pi_3 }{b} = \pi_1 $ and
$ \pi_1 \approx c_1 $ and $ \pi_2 \approx 0 $.

Alternatively, one can substitute out $x_1$ using the equality constraint $x_1 = t - bx_2$ and solve a one-dimensional optimisation problem in $x_2$ alone. This reduces the Gibbs sampler to sampling from a single truncated exponential distribution, which can significantly improve computational efficiency.

For simulation from the one-dimensional truncated exponentials we proceed as follows.
For the one-dimensional truncated exponentials we have:
density $ exp ( mx ) $ on $(a,t) $ where $ m>0$ has inverse cdf
$$
t+ \log \left ( \exp^{m(a-t)} + U(1- \exp^{m(a-t)} \right )/m
$$
where $ U \sim U(0,1)$.

\subsection{Example: Bayesian Portfolio Optimisation}

To illustrate the method's performance on a problem with uncertainty, consider the classical problem of maximising the expected utility of wealth from investing in a portfolio of risky assets.  In this case,  $G(\omega,x)=K-e^{-\gamma(r(\omega)^T x + r_f)}$, where $r(\omega)$ is an $n$-vector of the excess returns of the risky assets relative to a risk-free return $r_f$ and $K$ is chosen sufficiently large that the returns can be restricted so that $G(\omega,x)\ge 0$.  For $r$ normally distributed with mean $\mu$ and covariance $\Sigma$, the system can be solved analytically to obtain $x^*=\Sigma^{-1}\mu/\gamma$.

For the MCMC implementation, we introduce an auxiliary variable $w$ so that
$$
G(\omega,x)=\int_{-\log K/\gamma}^{r(\omega)^Tx + r_f}\gamma e^{-\gamma w}dw,
$$
i.e., so that $w$ is conditionally a truncated exponential random variable.  We then have
 $$
  \pi_J ( \underline{\omega}^J , x ) \propto \prod_{j=1}^J  \gamma e^{-\gamma w_j}1_{\{-\log K/\gamma\le w_j\le r_j^T x+r_f\}}e^{-(r-\mu)^T\Sigma^{-1}(r-\mu)/2}.
  $$
The MCMC iterations then iteratively draw $w_j$, returns $r_j$, and solutions $x^{(g)}$ as follows.
$$
w_j -\log(K)/\gamma \sim \mathcal{E}(\gamma,[0,\frac{\log(K)}{\gamma}+r_j^T x+r_f]) (\mbox{\rm\ truncated\ exponential\ with\ parameter\ } \gamma),$$
$$
r_j \sim \mathcal{N}((\mu,\Sigma)|r_j^Tx + r_f\ge w_j) (\mbox{\rm \ truncated\ normal)},$$
$$
x_k \sim \mathcal{U}([\max_{r_{jk}>0}\frac{w_j-r_{j\bar{k}}^T x_{\bar{k}}-r_f}{r_{jk}},\min_{r_{jk}<0}\frac{w_j-r_{j\bar{k}}^T x_{\bar{k}}-r_f}{r_{jk}}])  (\mbox{\rm \ uniform)},
$$
where $k=1,\ldots,n$ and subscript $\bar{k}$ denotes the vector of components other than $k$.

The sample-average approximation method tries to avoid a potentially exponential number of evaluations for $ \omega $
by simulating $ \underline{\omega} = ( \omega_1 , \ldots , \omega_N ) $ and approximating the criterion
function $ E [ G( \omega , x ) ] $ by the sample average
\begin{equation}\label{saa}
 \hat{G} (x) = \frac{1}{N} \sum_{i=1}^N G ( \omega_i , x ).
\end{equation}
The method then can use gradient-based methods to also estimate
the gradient using  MC samples  to find  $ {\rm argmax}_x  \hat{G} (x) $.  In the general, the solution $x^N$ to \eqref{saa}, while consistent, may be biased for finite $N$ with a bias that is difficult to detect without additional information.

\section{Stochastic Sampling Techniques}

The $MC^2$ methodology requires efficient algorithms for sampling from truncated distributions. In this section, we develop the necessary sampling techniques for multivariate exponentials, which arise in linear programming, and multivariate normals, which arise in quadratic programming.

\subsection{Truncated Sampling: Multivariate Exponentials}

The most direct approach to draw from a truncated multivariate exponential is to use one-at-a-time Gibbs sampling.
This can be slow to converge when there is an ``icicle" in the constraint region. First write $x = (x^+ , x^- )$ and
then transform the system so that all coordinates are positive $ x>0$. Also assume that the constraint is of the form $A x \leq b $ where
$ x $ is $ K \times 1 $ and $A$ is $ n \times K $ where typically $ n \geq K $.

The constraint $ A x \leq b $ can be written as a series of conditional constraints \citep{gelfand1992bayesian} where
$$
\sum_{j=1}^n a_{ij} x_j \leq b_i  \; {\rm implies} \;  a_{ik} x_k \leq b_i - \sum_{ j \neq k} a_{ij} x_j;
$$
therefore,
$ x_k \leq \min_i ( b_i - \sum_{ j \neq k} a_{ij} x_j ) / a_{ik} $ if the $a_{ik} > 0 $.

In some cases, we can use the factorisation $p( x_1, x_2 ) =p(x_1|x_2)p(x_2)$. For example, consider the joint distribution
$$
\exp \left ( - q^\prime x \right ) \mathbb{I} \left ( a_1 x_1 + a_2 x_2 \leq b \right )
$$
We can use the transformation $ z = a_1 x_1 $ and $ y - z = a_2 x_2 $; then, the distribution
$$
p( y | z ) \sim TExp_{ (z,b)} ( q_2 a_2^{-1} ).
$$
The marginal distribution $p(z)$ is then determined by
$$
p( z ) \propto e^{-z} \left ( 1 - e^{ - q_2 a_2^{-1} (b - z )} \right ) \; {\rm where} \; 0<z<b.
$$
This is a density tilted by a cdf, which can be sampled using the following result.

\begin{theorem} (Vaduva)
Generate $ X \sim f $ and $Y$ with cdf $ F $ until $ X \geq Y $.

This will gives samples from $p(x) =cf(x) F(x)$, where
$c$ gives the efficiency of the algorithm.
\end{theorem}

\paragraph{Conditioned to the Simplex}

The truncated multivariate exponential on the simplex $ x \in \mathcal{S}_{k-1} $ is defined by
$$
f(x) = d(\lambda)^{-1} \exp \left ( \sum_{j=1}^{k-1} - \lambda_j x_j \right ) \; \;
x = ( x_1 , \ldots , x_{k-1} )^\prime ; \; x_j \geq 0 , \sum_{j=1}^{k-1} x_j \leq 1
$$
\cite{kent2004simulation} calculate the normalisation constant for unequal and equal $\lambda $'s.
They write
$$
p_K ( \lambda )= d(\lambda) \prod_{j=1}^{k-1} \lambda_j
$$
where $ p_K ( \lambda ) = \mathbb{P} \left ( \sum_{i=1}^{k-1} X_i < 1 \right ) $ is the probability of the indept exponentials lying in the simplex.

If the $ \lambda $'s are all equal we get
$$
p_K ( \lambda ) = \left ( e^\lambda - \sum_{j=1}^{k-2} \frac{ \lambda^j}{j!} \right ) e^{- \lambda}.
$$

\paragraph{Equal $\lambda$'s}

If the $ \lambda_j = \lambda $ then this distribution can be easily sampled. Consider the transformation
$ x_j = y r_j $ where $ y =  \sum_{j=1}^{k-1} x_j $ is the ``size'' of $x$. After a change of variables,
$$
f( y,r) \propto y^{k-2} \exp \left ( - \lambda y \right ) \; , r \in \mathcal{S}_{k-2} \; , 0 \leq y \leq 1.
$$
We need to sample from the simplex and from a truncated gamma as follows.

\begin{itemize}
\item
It is straightforward to sample from the simplex. Take uniforms $ U_1 , \ldots , U_{k-2} $ on $[0,1]$ and let $ s_1^\prime , \ldots , s_{k-2}^\prime $
be the successive gaps in the order statistics, $ s_1^\prime = U_1 , s_j^\prime = U_{(j)} - U_{(j-1)} $ for $ j = 2 , \ldots , k-1$. See Method 2.
\item
Sampling from a truncated gamma distribution can be done with the ratios of uniforms method. See Method 3.
Simulating $ Y \sim \Gamma ( k-1 , \lambda ) $ on $[0,1]$ is equivalent to $ X = \lambda Y \sim  \Gamma ( k-1 , \lambda ) $ on $[0,\lambda]$.
Then if we let $ W - X - \min(k-1, \lambda) $ be a location shift. Let $f(w)$ be the pdf. Then we can define $ u_+ , v_- , v_+ $ (see Kent, p.56)
and generate $(U,V)$ uniformly in the rectangle
$$ D = [0 , u_+ ] \times [ v_- , v_+ ]
$$ and accept $ W = V/U$ if $ U < f^{\frac{1}{2}} ( V / U ) $.
\end{itemize}

In the general case of differing $ \lambda_j$'s. Kent proposes to use accept-reject on the simplex and calculates the rejection
probability using mgf methods. This takes a truncated independent exponential
$$
f_{TE} ( x) = b( \lambda ) f_E ( x ) \; , 0 < s_j < 1 , \; b(\lambda )^{-1} = \prod_{j=1}^{k-1} \left \{ 1 - \exp \left ( - \lambda_j \right ) \right \}
$$
and then conditions down to the simplex.
Method 1 applies for small values of the concentration parameter $\lambda$, Method 2 for large, and Method 3 for ranges between the others.

\subsubsection{Distribution of $ Y = \sum_{i=1}^{k-1} X_j $}

The Laplace transform of $ Y = \sum_{i=1}^{k-1} X_j $ where $ (X_j | \lambda_j ) \sim Exp( \lambda_j )$ is given by
$$
M (\phi) = \mathbb{E} \left ( \exp ( \phi Y ) \right ) = \prod_{j=1}^{k-1} \frac{1}{1 + \phi / \lambda_j }
$$
If the $ \lambda_j $ are all unequal this can be given by a partial fraction expansion
$$
M( \phi ) = \sum_{j=1}^{k-1} \frac{\lambda_j}{\lambda_j + \phi} \prod_{i=1, i \neq j}^{k-1} \frac{\lambda_i}{\lambda_i - \lambda_j}.
$$
Let $ a_j =  \prod_{i=1, i \neq j}^{k-1} \frac{\lambda_i}{\lambda_i - \lambda_j} $ where the $ \lambda $'s have been ordered.
This can be inverted to give the cdf of $ Y = \sum_{i=1}^{k-1} x_j $ as
$$
F( y) = \sum_{j=1}^{k-1} a_j ( 1 - e^{- \lambda_j y} ),
$$
Hence, the density for $Y$ is a weighted sum of exponentials, which is easy to simulate.

Setting $ y=1$ given the probability of lying in the simplex,
$$
p_T ( \lambda ) = \sum_{j=1}^{k-1} ( 1 - e^{- \lambda_j} ) \prod_{i=1, i \neq j}^{k-1} \frac{\lambda_i}{\lambda_i - \lambda_j}
$$
Hence, we can calculate the normalisation constant $d(\lambda)$.

\subsection{Exponential Slice Sampling}

An important enhancement to the basic Gibbs sampler is the introduction of auxiliary slice variables. Suppose that we wish to find $\max_{\mathcal{X}} f(x,y)$. We define the annealed distribution
$$
\pi_\kappa (x,y) = \exp \left ( \kappa f(x,y) \right ) / Z_\kappa \; \; {\rm where} \; \; (x,y) \in  \mathcal{X}
$$
If we introduce an auxilary exponential slice variable, so that the joint distribution is
$$
\pi( u , x , y ) \propto \kappa \exp ( \kappa u ) \mathbb{I} \left ( - \infty < u < f(x,y) \right ) \mathbb{I} \left ( (x,y) \in \mathcal{X} \right )
$$
Then, we have a truncated exponential for the slice variable
$$
\pi(u|x,y) = \frac{ \kappa \exp ( \kappa u ) }{ \exp ( \kappa f(x,y) ) } \mathbb{I} \left ( - \infty < u < f(x,y) \right )
$$
and a uniform for $(x,y)$ truncated via the bound $ u < f(x,y) $.
Marginalising out $u$ also gives us the appropriate marginal
$$
\pi_\kappa (x, y) = \exp \left ( \kappa f(x,y) \right ) \mathbb{I} \left ( (x,y) \in \mathcal{X} \right ) / Z_\kappa
$$
assuming this is a well-defined density.

This differs from the case of finding $ \min_{ \mathcal{X} } f(x,y) $ which requires a slice uniform on $(0, \exp ( - \kappa f(x,y) ) )$.

\subsection{Truncated Sampling: Multivariate Normal} 
We now turn to quadratic objectives, which arise in portfolio optimisation and quadratic programming. These lead to truncated multivariate normal distributions. Consider the problem of sampling from a truncated normal $\theta \sim N^T (\mu, \Sigma)$ on the constraint set $T = \{\theta : A\theta \leq b\}$. We denote this distribution by
\begin{equation}
    \theta \sim N^T (\mu, \Sigma) \quad \text{on} \quad \{\theta : A\theta \leq b\} 
\end{equation}
One approach is to transform the constraint set and let $\varphi = A\theta$ and rewriting the model as
\begin{equation}
    \varphi \sim N^T (A\mu, A\Sigma A^T) \quad \text{on} \quad \{\varphi : \varphi \leq b\} 
\end{equation}
This can lead to inefficient sampling due to the correlation structure of $\varphi$. Highly correlated variables will lead to poor Gibbs convergence properties. In order to do this we would have to use the conditionals
\begin{equation}
    \phi_k | \phi_{(-k)} \sim N (\mu^*_k, \sigma^2_{kk}) 
\end{equation}
\begin{equation}
    \mu^*_k = \mu_k + \Sigma^T_1 \Sigma^{-1}_{11}(y_1 - \mu_1) \quad \text{and} \quad \sigma^2_{kk} = \sigma_{kk} - \Sigma^T_1 \Sigma^{-1}_{11} \Sigma_1 
\end{equation}
An efficient alternative is to re-parameterise and use a variance stabilizing transformation. First find $Q$ such that $Q\Sigma Q' = I$ and set $\varphi = Q\theta$. Notice that the constraint set $S = \{Q\theta : A\theta \leq b\}$ transforms to $S = \{\varphi : D\varphi \leq b\}$ where $D = A Q^{-1}$. The problem becomes that of sampling
\begin{equation}
    \varphi \sim N^S (Q\mu, A\Sigma A^T) \quad \text{on} \quad \{\varphi : D\varphi \leq b\} 
\end{equation}
Let $\alpha = Q\mu$. We can now compute the complete conditionals and sample this joint distribution in a Gibbs fashion as follows
\begin{equation}
    \varphi_j | \varphi_{(-j)} \sim N^{S_j} (\alpha_j, 1) 
\end{equation}
where the conditional constraint set is given by
\begin{equation}
    S_j = \{\varphi_j : d_j\varphi_j \leq b - D_{(-j)}\varphi_{(-j)}\} 
\end{equation}
Since the constraint on y form a convex subset of $\Re^k$ the set $S_j$ can be determined as an interval of the form $[l_j, u_j]$, $(-\infty, u_j]$ or $[l_j, \infty)$ using the 1-dimensional constraints defined by $d_j\varphi_j \leq b - D_{(-j)}\varphi_{(-j)}$. Using the variance stabilizing transformation leads to a far more efficient algorithm. The algorithm can be summarised as follows:
\begin{enumerate}
    \item Initialize $\varphi^{(0)}$ satisfying the constraints $D \varphi \leq b$.
    \item For $t = 1, \ldots, T$:
    \begin{enumerate}
        \item For $j = 1, \ldots, k$:
        \begin{itemize}
            \item Calculate the bounds for $\varphi_j$ given the current values of $\varphi_{(-j)}$. For each constraint $i=1,\ldots,m$, we have
            \[
            d_{ij} \varphi_j \leq b_i - \sum_{l \neq j} d_{il} \varphi_l^{(current)}
            \]
            If $d_{ij} > 0$, this provides an upper bound; if $d_{ij} < 0$, a lower bound. The intersection of these intervals over all $i$ defines the conditional support $[L_j, R_j]$.
            \item Sample $\varphi_j^{(t)}$ from a truncated normal $N(\alpha_j, 1)$ constrained to $[L_j, R_j]$ using the inverse CDF method.
        \end{itemize}
    \end{enumerate}
    \item Transform back to the original parameters: $\theta^{(t)} = Q^{-1} \varphi^{(t)}$.
\end{enumerate}
This decorrelation step is crucial. In the original parameterization, the components of $\theta$ are correlated both due to $\Sigma$ and the constraints. The transformation removes the dependence due to $\Sigma$, leaving only the dependence due to the constraints, which is handled naturally by the truncated Gibbs steps.

\subsection{Stochastic Water-Filling Representation}

When implementing $MC^2$ algorithms with particle approximations, we need to devise a method of selecting a fixed number, say $N$, particles from the expanding set of particles to provide support economisation.
This problem reduces to that of approximating a discrete probability mass function, $ q = \{ q_j \}_{j=1}^M $
 with finite support by a \emph{stochastic} probability mass function, $ Q= \{ Q_j \}_{j=1}^M $ with
fewer support points. We need to determine $Q$ so that for a given value of $N<M$,

\begin{itemize}
\item $E(Q_j)=q_j$
\item The support of $Q$ has no more than $N$ points
\item $ \sum_{j=1}^M E( Q_j - q_j)^2 $ is minimised.
\end{itemize}

The unbiased condition, $ q_j = E(Q_j)$, lets us conclude  
$$
\sum_{j=1}^M f_j q_j = \sum_{j=1}^M f_j E ( Q_j ) = E_Q \left ( \sum_{j=1}^M f_j Q_j \right )
$$
for any functional $f$. We can use $ \frac{1}{N} \sum_{i=1}^N f_i Q_i $ to estimate the last term, that is averaging the functional over the current weights.

The water-filling solution is constructed as follows. Let $ N<M$ and $ \alpha$ be the unique root of $ N = \sum_{j=1}^M \min( \alpha q_j , 1) $. 
Then select particles with weights
$$
Q_j =  \frac{q_j}{p_j} \; {\rm with \; prob} \; \; p_j  = \min( \alpha q_j , 1 )
$$
and zero otherwise.
Equivalently, let $ p_j  = \min( \alpha q_j , 1 ) $ and select $Q_j$ such that
$$
Q_j = \left \{  \begin{array}{c}
q_j \; \; \; {\rm if} \; q_j \geq \frac{1}{\alpha} \; {\rm and} \;  p_j  = 1\\  
\frac{1}{\alpha} \; \; \; {\rm prob} \; \; p_j  = \min( \alpha q_j , 1 )
 \end{array} \right \}
$$
If $ q_j \geq 1/\alpha$ the weight stays the same (i.e. $Q_j = q_j$) and the particle is selected with probability one.
If $ q_j < 1/\alpha$, it is resampled with probability $ \alpha q_j$ and the new weight is set equal to $ Q_j = 1/\alpha$. 
This scheme satisfies the three required conditions described above. By definition of $\alpha$, we have $N$ particles left out of the $M$.

We apply this with $ q_j = p( \phi_{t+1}^{(j)} ) / \sum_{i=1}^N p( \phi^{(i)}_{t+1} ) $ at time $t+1$.  

The condition that with probability one at most $N$ of the $Q_j$'s are non-zero implies
that $Q_j=X_j$ with some probability $p_j$ and zero otherwise, where $ \sum_{j=1}^M p_j \leq N $. We can decompose the mean aquared error as
$$
E \left ( ( Q_j - q_j)^2 \right ) = p_j ( E(X_j) - q_j )^2 + p_j Var (X_j) + (1 - p_j) q_j^2
$$
Thus is minimised by taking $ X_j = c_j $, a constant. The unbiasedness condition $E(Q_j)=q_j$ implies that $ c_j = q_j /p_j $.
Then we have
$$
E \left ( ( Q_j - q_j)^2 \right ) = \sum_{j=1}^M q_j^2 \left ( p_j^{-1} -1 \right )
$$
Minmising this is equivalent to minimising
$$
\sum_{j=1}^M \frac{q_j^2}{p_j} \; \; {\rm subject \; to} \; \;  \sum_{j=1}^M p_j \leq N
$$
This is a standard constrained minimisation problem and the KKT condition gives
$$
p_j = \min ( \alpha q_j , 1 ) \; \; {\rm for} \; \; j=1,\ldots , M \; .
$$
Water-filling has the feature that it
``keeps the winners'' and ``re-sampler the losers''. We
minmise the weight mean squared error whilst retaining unbiaseness.
The $N$ particles will each having a possible $K$ ``children'' and the next mixture has $ M = NK $ particles which is then collapsed back to $N$ particles
and the process is repeated.

For the water-filling solution, the condition 
$$ 
\min( \alpha q_1 , 1 ) + \min( \alpha q_1 , 1 ) =N=1 
$$ 
implies that $ \alpha^{-1} = q_1 + q_2 $, neither particle is
above the bar and so the sequential collapsing mixture approach reduces to Barker's algorithm.

When $M \gg N$ with uniform weights $q_j = 1/M$, the condition $\sum_{j=1}^M \min(\alpha q_j, 1) = N$ implies $\alpha \approx 1/N$, and since $1/M < 1/N$, no particles exceed the threshold. In this regime, the water-filling approach offers no advantage over standard resampling. The method works best when there is an exponentially small number of high posterior probability solutions and one discovers them sequentially, with the ratio $M/N = K$ held fixed.

Several extensions of the basic water-filling scheme are noteworthy. First, the method can be applied sequentially, which prevents an exponentially large number of low-quality models from diluting the weight of good solutions. Second, the same argument extends to settings with latent variables. Suppose $q_j = \mathbb{E}(q_j(z))$ but the expectation cannot be evaluated directly. The procedure can be applied conditional on $z$, using the fact that
$$
\sum_{j=1}^M q_j(z) = 1 \; \forall z \; \; \text{implies} \; \; \sum_{j=1}^M \mathbb{E}_z(q_j(z)) = 1.
$$

The water-filling representation also applies to ratios of expectations, which arise in Bayes factor calculations:
$$
\frac{\mathbb{E}_q(f)}{\mathbb{E}_q(g)} = \sum_{j=1}^M \frac{q_j g_j}{\sum_{j=1}^M q_j g_j} \frac{f_j}{g_j}
 \approx \sum_{j=1}^N Q_j \frac{f_j}{g_j}.
$$
More generally, when $q_j = \mathbb{E}_v(q_j(v))$, one can simulate $v_j \sim p(v)$ and collapse the mixture over both indices $j$ and latent variables $v_j$, while preserving the unbiasedness property:
$$
\sum_{j=1}^M f_j q_j = \sum_{j=1}^M f_j \mathbb{E}_v(q_j(v)) = \mathbb{E}_{Q,v}\left(\sum_{j=1}^M f_j Q_j(v_j)\right).
$$

Finally, the method can be iterated using the law of iterated expectations. Writing $Q_j = \mathbb{E}(\tilde{Q}_j)$, we have
$$
\sum_{j=1}^M f_j q_j = \sum_{j=1}^M f_j \mathbb{E}(Q_j) = \mathbb{E}_Q\left(\sum_{j=1}^M f_j Q_j\right) = \mathbb{E}_Q\left(\sum_{j=1}^M f_j \mathbb{E}(\tilde{Q}_j)\right),
$$
which allows sequential refinement of the particle approximation.

\section{Applications}

\subsection{Example: Farmer Problem}

We demonstrate our methodology on the classical farmer problem from stochastic programming. This example is based on the crop allocation problem in \cite{birge1997introduction} for the farmer's decision of an initial storage decision $x$ and then subsequent allocation decisions $y_1$ and $y_2$ that depend on the realization of the random outcome.  The overall problem is to maximize profit of $-k x + E[143 y_1 + 60 y_2]$ subject to storage $y_1+y_2 \leq x$ and investment constraints,
$$
110 y_1 + 30 y_2 \leq \omega_1 \; \; {\rm and} \; \; 120 y_1 + 210 y_2 \leq \omega_2,
$$
with $y_1,y_2 \geq 0$.

We can directly use conditional exponential sampling or introduce a further exponential slice variable and then sample
from conditional uniforms.  In the former, we have conditionals
$$
p( y_1 | y_2 ) \propto \exp \left ( 143 \kappa y_1 \right ) \mathbb{I} \left ( 0 \leq y_1 \leq \min ( (\omega_1-30y_2)/110 , (\omega_2-210y)/120 , x-y_2) \right )
$$
$$
p( y_2 | y_1 ) \propto \exp \left ( 60 \kappa y_2 \right ) \mathbb{I} \left ( 0 \leq y_2 \leq \min ( (\omega_1-110y_1)/30 , (\omega_2-120y_1)/210 , x-y_1) \right )
$$
If we use the auxiliary slice variable we have the extra constraint $ u \leq 143 y_1 + 60 y_2 $.  Then we sample three conditionals
$$
p( u | y_1, y_2 ) = \kappa \exp ( \kappa u ) \mathbb{I} \left ( - \infty < u < 143 y_1 + 60 y_2 \right )
$$
\begin{multline*}
p( y_1 | x,y_2 ) \propto \mathbb{I} \left ( \max( 0 , ( u - 60 y_2)/143 ) \leq y_1 \right. \\ 
\left. \leq \min ( (\omega_1-30y_2)/110 , (\omega_2-210y_2)/120 , x-y_2) \right )
\end{multline*}
\begin{multline*}
p( y_2 | x,y_1 ) \propto \mathbb{I} \left ( \max ( 0 , ( u - 143 y_1 )/60 ) \leq y_2 \right. \\
\left. \leq \min ( (\omega_1-110y_1)/30 , (\omega_2-120y_1)/210 , x-y_1) \right )
\end{multline*}
The latter might be able to escape from local modes.

 Figures 2 and 3 present the results of these iterations for fixed $x=75$, $\omega_1=4000$, and $\omega_2=15000$. 

For an example, we make $J=20$ copies of the recourse variables and then follow the MCMC iterations over $x$, $y_2^j$, and auxiliary variables $u^j$.  For different values of $k$, with $\omega_1\sim \mathcal{U}(3000,500)$ and $\omega_2\sim \mathcal{U}(10000,20000)$, we have the objective values as a function of $x$ as given in Figure 4.  The histogram results for the last 2500 MCMC iterations of 5000 iterations appear in Figure 5.  The modal interval  again contains (or nearly contains) $x^*$ in each case.
\begin{figure}[hbp]
\includegraphics[height=6in,width=\textwidth]{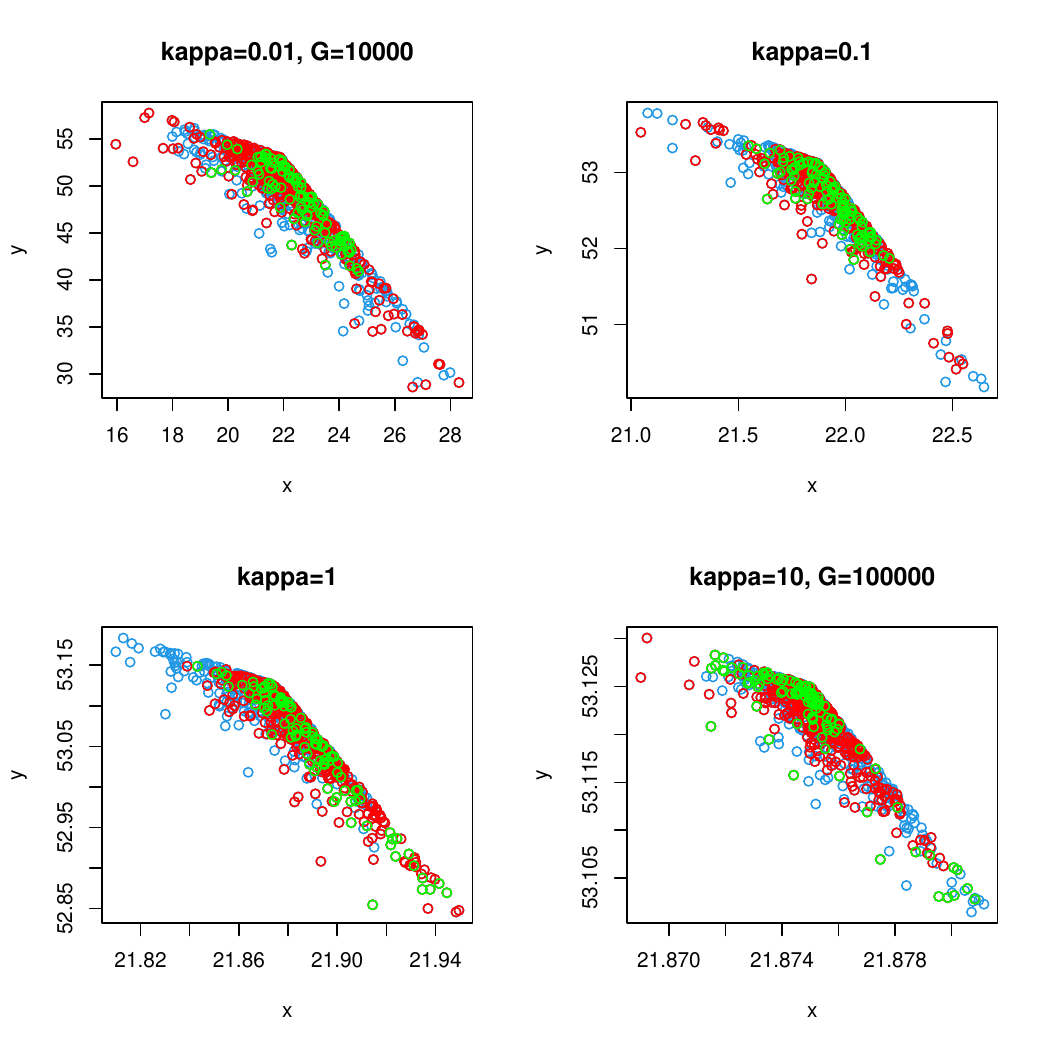}
\caption{Gibbs Output.}
\end{figure}

We need to calculate the normalising constant $Z(z,\omega )$.
The constraint region $ (y_1,y_2) \in \mathcal{X} (\omega , x ) $ is given by
$$
y_1,y_2 >0 , \; y_1+y_2 \leq x , \; 110 y_1 + 30 y_2 \leq \omega_1 , \; 120y_1 + 210 y_2 \leq \omega_2
$$
We will also assume that
$$
0 < x < 100 \; , \; 3000 \leq \omega_1 \leq 5000 \; , \; 10,000 < \omega_2 < 20000 \; .
$$
Given $x$ and $(\omega_1 , \omega_2)$, we can find the optimal $ x^\star $ as we know
$$
0 < y_1 < \min \left ( x -y_2 , ( \omega_1 - 30 y_2) /110 , ( \omega_2 - 210y_2 )/120 \right )
$$
For the $y$ variable, we have marginally $ 0<y_2< \min ( x , \omega_1 / 30 , \omega_2 / 210 ) $.
Given that $ x< 100 $ and $ \omega_1 > 3000 $, this reduces to $ 0<y_2< \min ( x , \omega_2 / 210 ) $.

Hence, we can substitute out and consider the nonlinear criteria function
$$
\mathbb{E}_\omega \max_y \left ( 60 y_2
 + 143 \min \left (x -y_2 , ( \omega_1 - 30 y_2) /110 , ( \omega_2 - 210y_2 )/120 \right ) \right )
$$

This leads to a marginal annealed distribution
where we use the Pincus result for nonlinear functionals.

We can therefore anneal only the distribution of $ (y_2 |\omega , x ) $ with density
$$
p_\kappa ( y_2 | \omega , x ) = \exp \left ( \kappa (  60 y_2
 + 143 \min \left ( x -y_2 , ( \omega_1 - 30 y_2) /110 , ( \omega_2 - 210y_2 )/120 \right ))  \right )
 / Z_\kappa (\omega , x )
$$
defined over the region $ \mathbb{I} \left ( 0 < y_2 < \min ( x , \omega_2 / 210 ) \right ) $.

This is a piecewise exponential distribution.
We still need to be able to either calculate the normalisation constant $  Z_\kappa (\omega , x ) $ explicitly, which follows from integrating the piecewise exponential distribution. An alternative is to directly sample from this distribution.
In this case the normalisation constant cancels in the Metropolis acceptance probability calculation.

%
%
%
%

\section{Discussion}

We have developed $MC^2$ algorithms for solving mixed integer and linear programming problems by reformulating them as simulation problems from Boltzmann distributions. The key insight is that constrained optimisation can be expressed as sampling from appropriately truncated distributions, namely multivariate exponentials for linear programs and multivariate Gaussians for quadratic objectives. As the annealing parameter $\kappa$ increases, these distributions concentrate on the optimal solutions.

The approach has several advantages over traditional optimisation methods. First, it provides a unified framework for handling both discrete and continuous decision variables, as well as linear and nonlinear constraints. Second, the MCMC framework naturally handles uncertainty in the problem parameters, making it well-suited for stochastic programming applications where one must simultaneously compute expectations and optimise. Third, the introduction of auxiliary slice variables enables the sampler to traverse multi-modal objective landscapes more effectively than standard simulated annealing.

Our methodology relies on efficient algorithms for sampling from truncated distributions. For truncated multivariate exponentials, the work of \cite{kent2004simulation} provides methods that work well across different parameter regimes. For truncated multivariate normals, variance-stabilising transformations following \cite{gelfand1992bayesian} lead to efficient Gibbs samplers. The stochastic water-filling representation offers a principled approach to particle economisation when the number of scenarios becomes large.

Several directions for future research merit attention. The computational complexity of our MCMC algorithms warrants further study, particularly in identifying problem structures that lead to polynomial-time mixing. Recent work on counting solutions to combinatorial problems via MCMC \citep{polson2024counting} suggests that certain problem classes may admit efficient algorithms. Additionally, combining our framework with variance reduction techniques such as importance sampling or control variates could improve the efficiency of expected value estimation in stochastic programming. Finally, extending the methodology to multi-stage stochastic programs with more than two stages, where the curse of dimensionality is most severe, remains an important challenge.

\bibliographystyle{plainnat}
\bibliography{SampleOpt}

\appendix
\section{Stochastic Programming Formulations}\label{app:stochastic}

\subsection{One-Stage Stochastic Programming}

The set-up for the one-stage problem is as follows. We seek $x^{\ast}$ such that
\be\label{onestage}
 E_{\omega} \left [ G( \omega , x^* ) \right ]=\max_x
E_{\omega} \left [ G( \omega , x ) \right ] ,
\ee

where $ \omega \sim p( \omega ) $ for $ G( \cdot , \cdot )$ given.
Let $G(x) = E_{\omega} \left [ G( \omega , x ) \right ] $. Assuming $G > 0 $ and
$$
\int E_\omega \left [ G( \omega , x ) \right ] d \mu(x) < \infty
$$
for some measure $ \mu( dx )$ to avoid singularities for $ x^{\ast} $. Let
$$
  \pi_J ( \underline{\omega}^J , x ) \propto \prod_{j=1}^J G ( \omega_j , x ) p( \omega_j ).
$$
The marginal $ \pi_J ( x ) \propto
E_\omega \left [ G( \omega , x ) \right ]^J $ has its mode at $x^*$ and  collapses on $ x^{\ast} $ as $J\uparrow \infty$ as
required (see \cite{pincus1968closed}).

The MCMC conditionals can then be found as:
$$
\pi_J ( \omega_j | x ) \propto G ( \omega_j , x ) p( \omega_j )
$$
and
$$
\pi_J ( x | \underline{\omega}^J ) \propto \prod_{j=1}^J G ( \omega_j , x ) p( \omega_j ),
$$
which can be sampled through Gibbs sampling or via Metropolis--Hastings algorithms. The result is a Markov chain with samples
$  \{ \underline{\omega}^{J,(h)} , x^{(h)} \}_{h=1}^H$.

The key property is the ability to simulate the $\omega_j$'s from a density that
depends on current state in the chain $x^{(h)}$.
Convergence is achieved with $ \{ \underline{\omega}^{J,(H)} , x^{(H)} \}
\rightarrow \pi_J (  \underline{\omega}^{J} , x ) $ and
$   x^{(H)} \rightarrow x^{\ast} $ in mode (and expectation as $J\to \infty$).

This approach requires no optimization step as in methods based on a sample--average approximation (see \cite{birge1997introduction}) and does not require the setting of step-length parameters as in
approaches based on stochastic approximation (see \cite{robbins1951stochastic} and \cite{rubinstein1999crossentropy}). By also not requiring properties such as convexity, the MCMC method allows for a wide range of objectives (such as black-box functions) that are not amenable to solution with other methods.

\subsection{Two-Stage Stochastic Programming}

The one-stage problem \eqref{onestage} can be extended to dynamic environments in which multiple actions take place sequentially as uncertainty is gradually resolved.  A simple version is the two-stage (linear) stochastic program with (fixed) recourse to find $ x^{\ast} $ to solve
$$
\min_{x \in S} c^T x + E_{\omega} \left [ Q ( x , \omega  ) \right ]
$$
where $ S = \{ Ax = b, x\ge 0\} $ and
$$
 Q(x, \omega) = \min_{y(\omega) \geq 0}
\{ q^{T}(\omega) y \| W y = h(\omega) - T(\omega) x \},
$$
$ y(\omega)$ is known as the recourse decision given  the realization of $\omega$.
See \cite{shapiro1996simulationbased} and \cite{birge1997introduction} for Monte Carlo solutions, and \cite{glasserman2003monte} for financial applications of Monte Carlo methods. We can convert a two-stage problem into a simulation problem that resembles the one-stage formulation.

The key observation is that, by duality, we can write the recourse function as $Q(x, \omega) = G(h(\omega) - T(\omega)x)$, where
$$
G(h - Tx) = \max_{\xi} \{(h - Tx)^T \xi : W^T \xi \leq q\}.
$$
MCMC simulation is far easier to apply with an inequality constraint $W^T \xi \leq q$ than with an equality constraint $Wy = h - Tx$. With this dual formulation, the two-stage problem becomes
$$
\max_{x \in S} \mathbb{E}_{\omega} \left[\max_{\xi : W^T \xi \leq q(\omega)} \{(h(\omega) - T(\omega)x)^T \xi + c^T x\}\right].
$$

For the two-stage stochastic program, the LP recourse problem can be replaced by an expectation under the annealed distribution
$$
p_\kappa ( \pi | \omega , x , q , W ) = \exp \left ( - \kappa \pi ( T(\omega) x - h(\omega ) ) \right )
\mathbb{I} \left ( W^\prime \pi \leq q \right ) / Z_\kappa (\omega.x)
$$
where $ Z_\kappa $ is an appropriate normalisation constant. We can then sample from this multivariate distribution in a number
of ways. For an ``over-determined" $W$ using one-at-a-time Gibbs leads to $min/max$ constraints.

Overall, the MCMC method leads to the limit
$$
\mathbb{E}_{\pi | \omega,x,q,W } \left ( \pi^\prime ( T(\omega) x - h(\omega ) ) \right ) \rightarrow
( \pi^\star )^\prime ( T(\omega) x - h(\omega ) ) \; \; {\rm as} \;   \kappa \rightarrow \infty
$$
Therefore, we can instead use MCMC methods to solve
$$
\max_x \mathbb{E}_\omega \left ( - \max_{\pi : W^\prime \pi \leq q    } \; \pi^\prime (T(\omega) x - h(\omega )) \right )
 = \max_x \mathbb{E}_\omega \left ( - \mathbb{E}_{\pi | \omega,x,q,W } \left ( \pi^\prime ( T(\omega) x - h(\omega ) ) \right ) \right )
$$
For MCMC sampling, we define a joint distribution
$$
p( x ,\pi , \omega ) \propto \pi^\prime q \cdot p_\kappa ( \pi | \omega , x ) p( \omega )
$$
where $ p_\kappa $ is the annealed distribution. This has the appropriate marginal
$ p( x ) \propto \mathbb{E}_\omega \left \{ \mathbb{E}_{\pi | \omega,x } \left ( \pi^\prime q \right ) \right \} $.
We can also introduce a further slice variable $u$ to deal with the objective function, where the slice variable satisfies $0 < u < \pi^\prime q$.

Consider the augmented joint distribution
$$
p( x , u , \pi , \omega ) \propto \mathbb{I} \left ( 0 < u < \pi^\prime q \right ) \cdot p_\kappa ( \pi | \omega , x ) p( \omega )
$$
with a uniform measure on $x$.

The conditional for the $\pi$ generation essentially adds another constraint
$$
\mathbb{I} \left ( 0 < u < \pi^\prime q \right )  \; \; {\rm and} \; \;
  \mathbb{I} \left ( W^\prime \pi \leq q \right ).
$$
We use one-at-a-time Gibbs to draw from the annealed
conditional $ p( \pi | \omega , x ) $ as a truncated multivariate exponential on this set of inequality constraints.

The uncertainty variable samples from an exponentially tilted distribution -- rather than the prior $p(\omega)$ in typical stochastic methods;
the conditional posterior is
$$
p( \omega | \pi , x ) \propto p( \omega ) \exp \left ( - \pi^\prime ( T(\omega )x - h( \omega ) ) \right )
$$
Finally, for the choice variable at the first stage,
$$
p( x | \pi , \omega ) \propto  \exp \left ( - \pi^\prime ( T(\omega )x - h( \omega ) ) \right ) \mu(x),
$$
where $\mu(x)$ is a uniform measure on the set $ \mathcal{X} $.

We can also add  $J$-copies over $ ( \omega , \pi )$ so that $p(x)$ again collapses on the optimum $x^*$.

\paragraph{Conditional Sampling}

\cite{gelfand1992bayesian} propose using Gibbs sampling to draw from the annealed distribution. This requires only the one-dimensional conditionals
$$
p_{\kappa} ( \xi_j | \xi_{(-j)} ,  a , W , \gamma ),
$$
where $\xi_{(-j)}$ denotes the vector of all variables except $\xi_j$.

If the inequality constraints $W^T \xi \leq \gamma$ can be written as $\xi_j \leq b_{(-j)}(W, \xi_{(-j)}, \gamma)$, then the one-dimensional conditionals are simply truncated exponentials:
$$
 \xi_j | \xi_{(-j)} ,  a , W , \gamma  \sim \frac{1}{ \kappa a_j }
 e^{ \kappa a_j ( \xi_j - b_{(-j)} )}  \; \; {\rm on} \; \;
  \xi_j \leq b_{ (-j)} ( W , \xi_{(-j)} , \gamma ).
$$
In the two-stage case, $a = a(\omega, x)$ and $\gamma = \gamma(\omega, x)$ depend on the random outcome and first-stage decision, leading to the conditional distribution
$$
p_{\kappa} ( \xi | \omega , x  ) \propto e^{ \kappa a^{\prime} ( \omega , x ) \xi }
\mathbb{I}_{\{ W^T \xi \leq \gamma ( \omega , x )\}},
$$
which makes explicit the conditioning on $(\omega, x)$.

The optimum is now a decision function $\xi^{\star}(\omega, x)$, which can be determined by simulating from the joint distribution
$$
p_{\kappa} ( \xi , \omega , x  ) = \frac{ e^{ \kappa a^{\prime} ( \omega , x ) \xi } }{ C( \omega , x) }
\mathbb{I}_{\{ W^T \xi \leq \gamma ( \omega , x )\}}
$$
for some reference measure $\mu(\omega, x)$ that ensures integrability and positivity. For large $\kappa$, plotting the $\xi$ draws versus $(\omega, x)$ reveals the decision function.

Here $C(\omega, x) = \int_{W^T \xi \leq \gamma(\omega, x)} e^{\kappa a^{\prime}(\omega, x) \xi} d\xi$ is typically not available in closed form. However, MCMC uses the conditionals
$$
p_{\kappa} ( \xi_j | \xi_{(-j)} , \omega , x )
\; \; {\rm and} \; \; p_{\kappa} ( \omega , x | \xi ),
$$
and the former does not require the normalisation constant---it is just a truncated exponential with parameters depending on $(\omega, x)$. The latter,
$$
p_{\kappa} ( \omega , x | \xi ) \propto
\frac{ e^{ \kappa a^{\prime} ( \omega , x ) \xi } }{ C( \omega , x) }
\mathbb{I}_{ W^T \xi \leq \gamma ( \omega , x )} \mu ( \omega , x ),
$$
does depend on $C(\omega, x)$. We can avoid direct evaluation by using the Metropolis algorithm and noting that, by the Clifford--Hammersley theorem, the ratio for any two candidate draws $(\omega, x)^{(g)}$ and $(\omega, x)^{(g+1)}$ can be computed as
$$
\frac{ p_{\kappa} ( \omega^{(g)} , x^{(g)} | \xi )}{ p_{\kappa} ( \omega^{(g+1)} , x^{(g+1)} | \xi )}
= \prod_{j=1}^k
\frac{ p_{\kappa} ( \xi_j | \xi_{(-j)} , \omega^{(g)} , x^{(g)} )}{ p_{\kappa}
(  \xi_j | \xi_{(-j)} , \omega^{(g+1)} , x^{(g+1)} )}
$$
in terms of the normalisation constants of the one-dimensional conditionals, which are known in closed form.

\subsection{Slice Sampling}

The intuition behind uniform slice sampling is simple. Suppose that we wish to sample from a possibly high dimensional un-normalised density
$ \pi(x) $. We do this by sampling uniformly from the region that lies under the density plot of $\pi$. This idea is formalised
by letting $u$ be an auxiliary ``slice-variable'' and defining a joint distribution $ \pi(x,u)$ that is uniform
on the set $ U = \{ ( x , u) : 0 < u < \pi(x) \} $.  Therefore, $ p ( x , u) = 1 / Z $ on $U$ and zero otherwise.
Here $ Z = \int_{\mathcal{X}} \pi(x ) d x $ is the appropriate normalisation constant.
The marginal is the desired normalised density as
$$
\pi(x) = \int_U \pi(x,u) d u = (1 / Z ) \int_0^{\pi(x)} d U = \pi(x)/Z  \; .
$$
We are then left with sampling from the uniform density on $U$. \cite{neal2000slice} provides a general slice algorithm. When it is straightforward to sample from the ``slice'' region defined by $u$, namely
$ \mathcal{S}_u = \{ x : u < \pi(x) \} $, then a simple Gibbs sampler which iterates between drawing a uniform $ (u|x) \sim Uni ( 0, \pi(x) ) $
and $ (x|u) \sim Uni_{ \mathcal{S}_u } ( x ) $ provides a Markov chain with the appropriate joint distribution $\pi(x,u)$ and hence marginal $\pi(x)/Z$.

This is a special case of the so-called Swendsen--Wang algorithm \citep{edwards1988generalization}. Suppose that we wish to sample from a density that is a product of functions: $p(x) = \pi_1 (x ) \ldots \pi_K (x) / Z_K $. Then we introduce a set of $K$ auxiliary uniform slice variables $ (u_1 , \ldots , u_K )$ and a joint $(x, u_1 , \ldots, u_K )$ that is uniform on the ``slice'' region:
$$
 \mathcal{S}_u = \{ x : u_i < \pi(x) \; \; \forall i \; \}
$$
Then we sample in a Gibbs fashion, from the complete conditionals
$$
(u_i | x ) \sim Uni ( 0, \pi_i(x) ) \; {\rm for} \;  i = 1 , \ldots , K  \; \; {\rm  and} \;  \; (x|u) \sim Uni_{ \mathcal{S}_u } ( x ) \; .
$$
In the case where the objective function of interest is additive and
$ f(x) = \sum_{i=1}^K f_i (x) $, we have a more structured joint distribution given by
$$
\pi_\kappa ( x ) = \exp \left ( - \kappa f (x) \right )/ Z_\kappa  = \exp \left ( - \kappa \sum_{i=1}^K f_i (x) \right ) / Z_\kappa \;.
$$
When constructing our Markov chain, we also wish to
perform a sensitivity analysis to the cooling parameter $\kappa$. Therefore, we modify the standard uniform slice sampler to
the exponential slice sampler as follows. Define an auxiliary latent
random variable $ (y | \kappa ) \sim Exp ( \kappa )$ and a joint distribution:
$$
\pi_\kappa (  x , y ) \propto p(y|\kappa) \mathbb{I}\left ( 0 \leq y \leq f(x) \right ).
$$
The auxiliary variable is exponential with mean $1/\kappa$ and hence $ p( y)= \kappa e^{-\kappa y} $.
To check that this has the appropriate marginal distribution, we have
$$
\pi_\kappa ( x , y ) \propto e^{ - \kappa y } \mathbb{I}\left ( 0 \leq y \leq f(x) \right )
$$
Integrating out $y$,
$$
\pi( x )  \propto \int_{f(x)}^\infty \kappa e^{-\kappa y} d y = \exp \left ( - \kappa f (x) \right )
$$
as required. This leads to a simple Markov chain based on its complete conditionals and a Gibbs sampler of the form:
\begin{align*}
\pi( x|y ) & \propto \mathbb{I}\left ( f(x) \geq y \right ) =  \mathbb{I}\left ( x \in f^{-1} ( y ) \right )  \\
\pi( y|x ) & \propto e^{-\kappa y} \mathbb{I}\left ( y \leq f(x) \right ) \; .
\end{align*}
For ease of implementation, we have to be able to compute the slice set and the  set-theoretic inverse $f^{-1}$.
We also need to be able to sample from a truncated version of $f$ and from a truncated exponential.

There are a number of extensions of this algorithm. First, suppose that we wish to sample from
$ \pi_\kappa (x ) = g(x) \exp \left ( - \kappa f(x) \right ) / Z_\kappa $ where $ g(x) $ and its truncated counterpart are straightforward to sample \citep{devroye1986nonuniform}.
Then we can slice the last term and consider the augmented joint distribution:
$$
 \pi_\kappa (x, y ) = g(x) e^{ - \kappa y } \mathbb{I}\left ( 0 \leq y \leq f(x) \right ).
$$
In a similar fashion this leads to a simple Gibbs sampler with the only difference being that we have to draw from $g(x)$ on the conditioned slice set given $y$.

Second, by introducing multiple exponential slice variables, as in the Swendsen--Wang algorithm, we extend this to densities of the form:
$$
\pi_\kappa ( x ) = \exp \left ( - \kappa \sum_{i=1}^K f_i (x) \right ) / Z_\kappa
$$
The advantage of our approach is that it works seamlessly for large values of $K$. Hence we can deal efficiently with multi-modal functions.

Specifically,  define multiple independent exponential slice variables $y_1 , \ldots , y_K $ and a joint distribution:
$$
\pi(  x , y_1 , \ldots , y_k ) = \exp \left ( - \kappa \sum_{i=1}^K y_i \right ) \prod_{i=1}^K \mathbb{I}\left ( 0 \leq y_i \leq f_i(x) \right ) / Z_\kappa.
$$
For the $ p( x | y_1 , \ldots , y_K )$ conditional we now need to sample from
$$
\pi( x|y )  \sim Uni \left \{   f_i(x) \geq y_i \; \forall i \;  = \cup_{i=1}^K \left ( x_i \in f^{-1}_i ( y_i ) \right ) \right . \}
$$
\cite{neal2000slice} for a general approach for dealing with sets of this form.

Finally, we can use the first coordinate of the draws from the joint distribution as a sample from the marginal dustribution
$\pi_\kappa (x)$ of interest.
So far we have constructed a Markov chain that generates a sequence of draws $\left ( x^{(n)} , y_1^{(n)} , \ldots,  y_k^{(n)} \right ) $
which converges in distribution as $ n \rightarrow \infty $ to a draw $( x , y_1 , \ldots , y_k) \sim \pi $ from the joint distribution of interest.
We write
$ \left ( x^{(n)} , y_1^{(n)} , \ldots , y_k^{(n)} \right ) \stackrel{D}{=} ( x , y_1,\ldots ,y_k ) \sim \pi $.
Given standard properties of weak convergence (or convergence in distribution), we have weak convergence for functionals $F: \Re^K
\rightarrow \Re^p $ where $ p \leq K $ and we have
$$
F \left ( x^{(n)} , y_1^{(n)} , \ldots , y_k^{(n)} \right ) \stackrel{D}{=} F ( x , y_1,\ldots , y_k ) \; .
$$
We can use this to estimate means $ \mathbb{E}_\pi \left ( g ( x ) \right ) $ of posterior functionals of interest by appealing to the ergodic theorem and using a delayed average mean estimator $ \frac{1}{N} \sum_{n=1}^N f( X^{(n)} ) $
along the dependent draws of the chain.

This approach can also be used to perform marginal density estimation. Suppose that we require an estimator $ \hat{\pi}(x) $ of
the given $ F ( x, y_1,\ldots , y_k ) = x $. Therefore, we have $ x^{(n)} \stackrel{D}{=}  x $ as
$ n \rightarrow \infty $.  We can  average along the chain to obtain the histogram density estimator
of the marginal distribution:
$$
\hat{\pi} ( x )= \frac{1}{N} \sum_{n=1}^N \delta_{ x^{(n)} } ( x ),
$$
where $ \delta_x ( \cdot) $ is the Dirac measure at point $x$.

\end{document}